\documentclass[a4paper, amsfonts, amssymb, amsmath, reprint, preprintnumbers, showkeys, nofootinbib, twoside, floatfix]{revtex4-2}
\usepackage[dvipsnames]{xcolor}
\usepackage{float}
\usepackage{dsfont}
\usepackage[english]{babel}
\usepackage[utf8]{inputenc}
\usepackage[colorinlistoftodos, color=green!40, prependcaption]{todonotes}
\usepackage[pdftex, pdftitle={Article}, pdfauthor={Author}]{hyperref} 

\begin{document}
\preprint{IMSC/2025/05}
\title{Topological Strings in SU(3) Gauge Theory at Finite Temperature}

\author{Sanatan Digal}
\email{digal@imsc.res.in}
\affiliation{The Institute of Mathematical Sciences, Chennai, 600113, India}
\affiliation{Homi Bhabha National Institute, Training School Complex, Anushaktinagar, Mumbai 400094, India}

\author{Vinod Mamale}
\email{vinod.mamale@wits.ac.za}
\affiliation{The Institute of Mathematical Sciences, Chennai, 600113, India\\
\&\\
National Institute for Theoretical and Computational Sciences,
School of Physics, and Mandelstam Institute for Theoretical Physics,
University of the Witwatersrand, Johannesburg, Wits 2050, South Africa}

\author{Sumit Shaw}
\email{sumitshaw@imsc.res.in}
\affiliation{The Institute of Mathematical Sciences, Chennai, 600113, India}
\affiliation{Homi Bhabha National Institute, Training School Complex, Anushaktinagar, Mumbai 400094, India}

\begin{abstract}

We investigate string configurations in the deconfined phase of SU(3) gauge theory, which arise from the spontaneous breaking of the $Z_3$ center symmetry. These configurations form at the junctions of domain walls of the theory. The complex phase of the Polyakov loop changes by multiples of $2\pi$ on large spatial loops around the string, rendering them topologically stable. Using the Monte Carlo simulations of the partition function,  we compute the free energy associated with these configurations. The simulations are performed on lattices with spatial dimensions $N_{x,y}=60, N_z=4$, and temporal extent $N_\tau=2,4$. Our results show that the free energy of the $Z_3-$strings is dominated by the domain walls. Further near the transition point, thermal fluctuations cause the decay of domain walls as well as the $Z_3$ strings into confined-deconfined interfaces.

\end{abstract}
\maketitle
\section{Introduction}
The non-perturbative study of Quantum Chromodynamics (QCD), plays a crucial
role in understanding matter under extreme conditions. In particular, it provides key insights into the phase diagram and the transition from the hadronic phase to the quark-gluon plasma (QGP). The thermodynamics of QCD remains an active field of research, driven by relativistic heavy-ion collision(HIC) experiments that probe different regions of the phase diagram. While the chiral and heavy-quark limits do not correspond to physical mass parameters, investigating these limits has significantly enhanced our understanding on the role and possible effects of chiral symmetry breaking and the confinement-deconfinement aspects in realistic QCD \cite{Satz:1982fu, Kallman:1982yc, Celik:1983nc, DeGrand:1986tp, Letessier:1994wm, Kalashnikov:1996nf, Bali:1993fb, Caselle:1994df, Belyaev:1991gh, Lenz:1998eu, Gazdzicki:1999ej, Baldo:2003id, Fiore:2004pg, Nefediev:2009kn, Greensite:2012dy, Sakai:1997ia, Borsanyi:2010cw, APE:1987ehd, Bali:2005fu, Cornwall:1982zn, Dubin:1994vn, Petreczky:2012rq, DeTar:2009ef, Baker:2024peg, Bicudo:2011hn}. 

QCD in the heavy-quark limit, i.e., pure $SU(3)$ gauge theory, has been extensively studied using lattice simulations, which offer a powerful approach to explore the non-perturbative regime. In finite-temperature studies of this theory\cite{Svetitsky:1982gs, Celik:1983wz, Boyd:1996bx, Iwasaki:1992ik, Lucini:2005vg}, it has been firmly established that the system undergoes a first-order confinement-deconfinement (CD) phase transition, at the critical temperature ($T_c$), between the hadronic phase and the QGP phase. The hadronic and the QGP phases can be interpreted in terms of glueballs and a thermalized medium of gluons, respectively. In pure $SU(N)$ gauge theories, the Polyakov loop expectation value($L$), which transforms non-trivially under the $Z(N)$ gauge transformations, serves as the order parameter for the CD phase transition. In the deconfined phase, the Polyakov loop, acquires a nonzero expectation value, i.e., $L\ne0$, leading to spontaneous breaking of the $Z(N)$ symmetry. Consequently there are $N$ degenerate states. For $N\ge 3$, the Polyakov loop is complex valued and these states are characterised by complex phases of the Polyakov loop, given by,
\begin{equation} 
	\theta_k={2\pi k\over N},~ k=0,1,..,N-1. 
\end{equation}
In contrast, for $N=2$, the Polyakov loop is real-valued, and the states are distinguished by its sign. These vacua of discrete states give rise to static domain walls (interfaces) that interpolate between a pair of vacuum sectors, i.e., $\theta_i$'s \cite{Mermin:1979rmp}.

The domain walls in Pure $SU(3)$ gauge theory, have been studied using lattice simulations, specifically
for $N_\tau = 2$ \cite{Kajantie:1990bu, Aoki:1993dm, Holland:1999mx} and for $N_\tau = 4$ \cite{Huang:1990jf}. A more recent
study on the calculation of interface tension for $N_\tau \geq 2$ has been conducted using different numerical algorithms
and employing the t’Hooft loop dual operator, with significant findings reported in
\cite{deForcrand:2005pb, deForcrand:2004jt, deforcrand:2005, Frei:1989es}. Details on the $2+1D$ simulation of the $Z_3$ interfaces
can be found in \cite{West:1996sb, West:1996ej, Korthals_Altes:1995}; these studies provide a comprehensive numerical
framework that captures the complex behaviour of these interfaces under various conditions. In earlier work, the
interface tension was calculated within a perturbative framework for $SU(N)$ gauge theories at finite temperatures, and this
was achieved by establishing a connection between the $Z(N)$ vacua and the instanton solutions of an effective action that
incorporates both classical and quantum fluctuations \cite{Bhattacharya:1992qb, Weiss:1980rj}.

\par
For $SU(N)$ gauge theory in the deconfined phase, the interface tension, $\alpha_{ij}$, of the domain walls depends on the
difference in the phases of the Polyakov loop, $|\theta_i-\theta_j|$, apart from temperature (T)
\cite{deForcrand:2004jt, Giovannangeli:2001bh}. Further, $\alpha_{ij}+\alpha_{jk} \ge \alpha_{ik}$, thus the interaction between these domain walls is attractive. The Polyakov loop takes on different thermal expectation values on either side of the domain walls. Its magnitude decreases as one approaches the center from either side of the domain wall, where it reaches a non-zero minimum value (see Section II). As a result, in a configuration of {\it multiple} domain walls, the phase of the Polyakov loop remains well-defined in most of the physical space. This suggests the possibility of an intriguing configuration in the deconfined phase, at the intersection of all the $N$ vacua. The domain walls can be oriented such that, the topological considerations impose a line-like intersection \cite{Layek:2005fn, Balachandran:2001qn}, where the value of the Polyakov loop $L(\vec{x})$ becomes exactly zero. As $L(\vec{x})$ becomes zero, locally there is restoration of $Z_N$ symmetry and confining behaviour in an overall deconfined environment. Thus, at high temperatures, the vacuum structure is such that the multiple degenerate $Z(N)$ states meet to form a defect whose core resembles the confined phase. This picture has been further elucidated by studies that employ lattice simulations and effective models to explore nonperturbative features of the quark-gluon plasma. In particular, for $N=3$, string configurations were investigated  \cite{Layek:2005fn} using an effective potential for the Polyakov loop with $Z_3-$symmetry \cite{Pisarski:2000eq}, providing valuable insights into how the order parameter deviates sharply from the value corresponding to the deconfined phase, due to the underlying topological structure.  These results, which are highly dependent on the effective potential, can only be validated through first-principles calculations, such as lattice simulations of the thermal partition function. 

In this paper, we explore the topological string configurations in $SU(3)$ gauge theory using lattice simulations, investigating their intricate role in elucidating the non-perturbative phenomena inherent in QCD. The free energy for these configurations is computed by integrating the action difference between systems with and without the string, a standard technique in statistical physics. In our study we also compute the interface tension, which is in good agreement with previous results \cite{Kajantie:1990bu}, reinforcing the consistency of our approach with earlier studies and highlighting the relevance of these methods in probing non-trivial topological configurations in gauge theories.

The $Z_3$ topological string studied in this paper and flux-tubes in QCD are fundamentally different. In the case of the $Z_3$ string, it is a deconfined phase everywhere except at the core. The situation reverses for flux-tubes, i.e., the phase outside is confined and is deconfined inside. The QCD strings that emerge in the confining phase of quantum chromodynamics (QCD), where they manifest as flux tubes linking quark-antiquark pairs or forming glueball states within the framework of pure gluonic theory—a perspective that is continually refined by numerical and analytical studies \cite{Hansson:1982dv, APE:1987ehd}. Unlike the QCD strings, which are prone to breaking via the spontaneous creation of quark-antiquark pairs \cite{Detar:1998qa, Bali:1992ab, Nefediev:2009kn, Johnson:2000qz}, the strings we examine are topologically stable, which allows them to persist and interact in the deconfined phase.

In this study, the effects of the dynamical quarks on the string configurations are neglected. However, previous research provides substantial evidence suggesting that quarks can influence such configurations. Both perturbative and mean-field studies indicate that the presence of quarks explicitly breaks the $Z_3$ symmetry \cite{Green:1983sd,Hasenfratz:1983ce,Satz:1985js,Belyaev:1991np}. The explicit breaking weakens the CD transition from a pure gauge first order transition to a crossover with decrease in the quark masses in the heavy quark regime \cite{Green:1983sd,Hasenfratz:1983ce,Satz:1985js}.  The explicit breaking lifts the degeneracy between the vacua. Since the three domain walls, in other words three vacua, join to form the string configuration, the lifting of degeneracy between the $Z_3$ vacua renders the string non-static. It will move towards the region of meta-stable states so as to reduce the free energy of the system \cite{Gupta:2011ag}. In the presence of dynamical quarks, near $T_c$, only the $\theta=0$ state is observed in the deconfined phase. The states with $\theta=\pm 2\pi/3$ appear as metastable states only above a certain temperature $T_m$ \cite{Deka:2012}. This suggests that the string configuration will become unstable and "melt" as the temperature drops below $T_m$. Analysis within the framework of the PNJL model at zero baryon chemical potential shows that if metastable states were to form in heavy-ion collision(HIC) experiments, they will survive until the temperature falls below $T_m$ \cite{Biswal:2019xju}. This suggests that string configurations may potentially affect the dynamics of QGP above $T_m$ in HIC. 
 
This paper is organised as follows. In Section II, we will briefly describe the $Z_3$ symmetry in the continuum and emergence of the string solution using the Polyakov loop effective potential.  The lattice gauge action and the lattice  Polyakov loop operator, emphasising its role, are presented in Section III.  Further, we describe the methodology for calculating the free energy via an indirect approach, since directly estimating the partition function is computationally challenging. Section IV outlines the numerical setup for studying string configurations and provides the technical details required to estimate the free energy of a string configuration. The conclusions are presented in Section V.
 
\section{$Z_3-$string in pure $SU(3)$ gauge theory}
\begin{figure}[t]
	\centering
	\includegraphics[width = 0.80\linewidth]{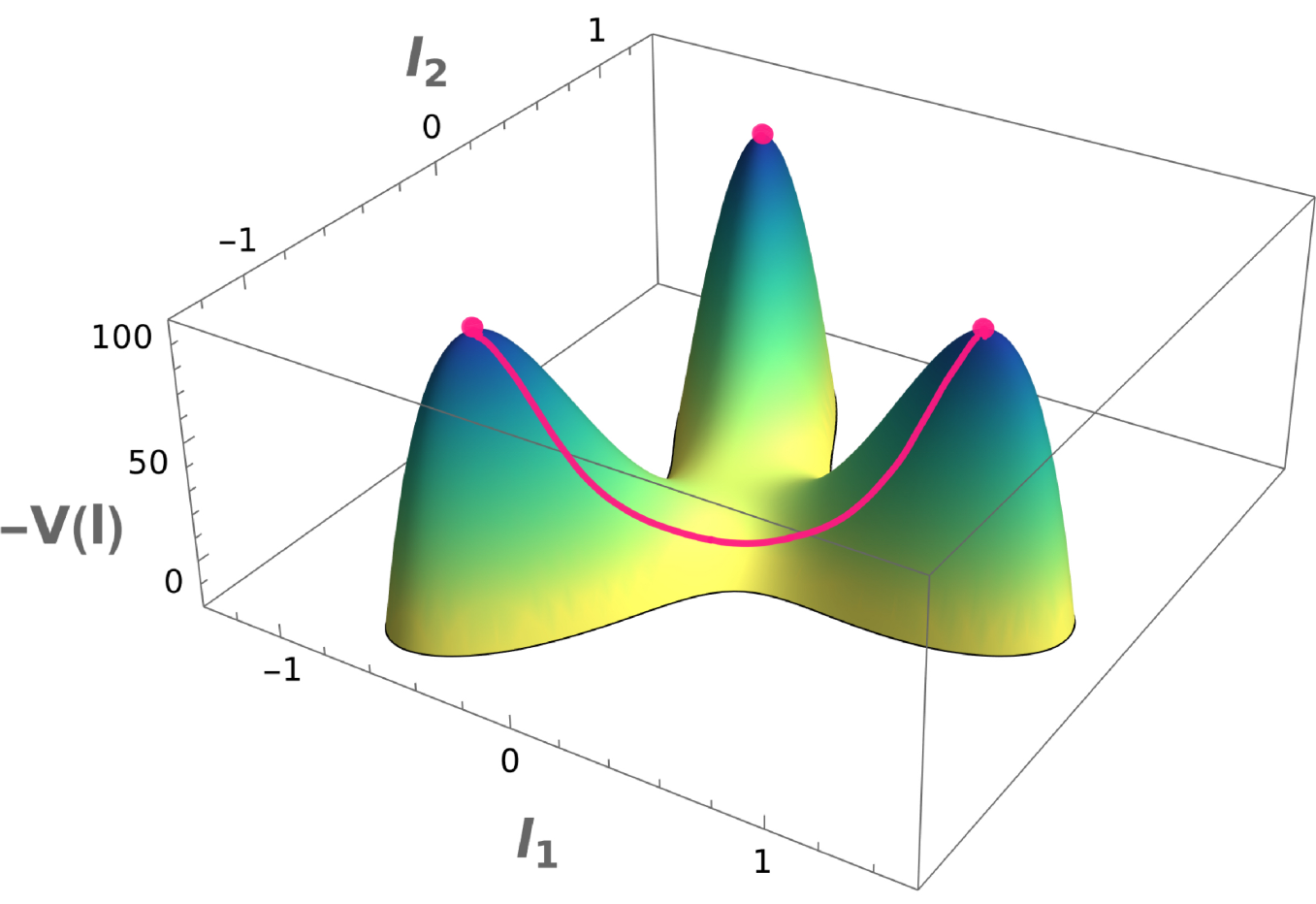}
	\caption{The inverted potential, $-V(L)$ vs $L$, in the complex $L-$plane for $T>T_c$.}
	\label{fig:epot2}
\end{figure}

In this section, we briefly discuss the $Z_3$ symmetry in pure $SU(3)$ gauge theory at finite temperature and how topological defects follow from the Polyakov loop effective potential. In pure $SU(3)$, the gauge fields $A_\mu$ at any point in space-time, are matrices in color space, i.e., $A_\mu=A^a_\mu T^a$, where $T^a, a=1,2...8$, are generators of $SU(3)$ \cite{Georgi:1999wka}. The partition function in the path-integral formulation is given by,
 \begin{eqnarray}
 {\cal Z} = \int \prod_\mu D A_\mu e^{-S_g}.
 \label{caction}
 \end{eqnarray}
 $S_g$ is the Euclidean gauge action, given by,
 \begin{eqnarray}
 S_g&=&{1 \over 4}\int_0^{1/T} d\tau \int d^3x F_{\mu\nu} F^{\mu\nu},\\
 F^{\mu\nu} &=& \partial_\mu A_\nu - \partial_\nu A_\mu + g \left[A_\mu,A_\nu\right].\nonumber
 \label{cgaction}
 \end{eqnarray}
 $T$ is the temperature and $g$ is the gauge coupling. The above action remains invariant under gauge transformations,
 $\Lambda(x)\equiv\Lambda({\bf x},\tau)$, i.e., 
\begin{eqnarray}
A_\mu(x) \rightarrow A^g_\mu(x)=\Lambda(x) A_\mu(x)\Lambda^{-1}(x) + i\Lambda(x) \partial_\mu \Lambda^{-1}(x).\nonumber
\end{eqnarray}
The path integration in Eq.\ref{caction} is carried out over gauge fields, $A_\mu(x)\equiv A_\mu({\bf x},\tau)$, that are periodic in the temporal
direction, i.e., $A_\mu({\bf x},0)=A_\mu({\bf x},1/T)$. This condition requires the gauge trasformations, $\Lambda(x)$, to be periodic in $\tau$, up to a factor $z$, i.e., 
\begin{eqnarray}
\Lambda({\bf x},0) = z \Lambda({\bf x}, 1/T), \nonumber
\label{gtr}
\end{eqnarray}
where $z\in Z_3$ and $Z_3$ is the centre of $SU(3)$. Consequently,
all the allowed gauge transformations are classified by the $Z_3$ group. Under these gauge transformations, the Polyakov loop($L({\bf x})$),
\begin{equation}
L({\bf x}) ={1 \over 3} {\rm Tr}\left\{ P\left(\exp\left[ ig\int_0^{1/T}d\tau A_0 ({\bf x}, \tau)\right]\right)\right\},
\end{equation}
transforms as $L({\bf x}) \rightarrow zL({\bf x})$, similar to how $Z_3$ spins transform. For temperatures $T\ge T_c$, in the deconfined phase, the thermal and the volume ($V_s$) average of the Polyakov loop,
\begin{eqnarray}
L(T)={1 \over {\cal Z}} \int \prod_\mu D A _\mu \left[{1 \over V_s}\int d^3x L({\bf x})\right] e^{-S_g},
\end{eqnarray}
acquires a non-zero value, which leads to the spontaneous breaking of the $Z_3$ symmetry.
 \begin{figure}[t]
\centering
\includegraphics[width = 0.80\linewidth]{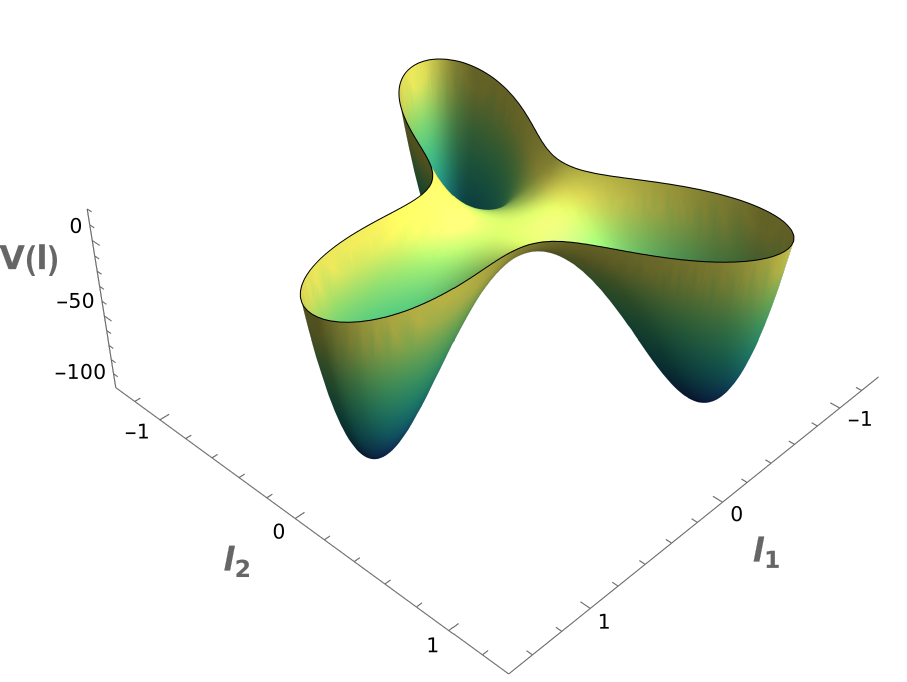}
\caption{$V(L)$ vs $L$, in the complex $L-$plane for $T>T_c$.}
\label{fig:epot}
\end{figure}
Consequently, the effective potential has three degenerate minima, as illustrated in Fig.\ref{fig:epot}. In the complex $L-$plane, the point
$L=0$ corresponds to a saddle point of the effective potential, while the minima occur for $|L|\ne 0$ with
the polar coordinate(phase) $\theta=0, 2\pi/3$, and $-2\pi/3$ \cite{Pisarski:2001pe, Weiss:1980rj}, which we denote by $L_0, L_1$ and $L_2$ respectively.
Given this, there are three possible domain walls, i.e., $L_{ij}$ interpolating $L_i$ and $L_j$. A static domain wall solution, on the $yz-$plane, is obtained by solving the following field equations,
\begin{eqnarray}
{d^2 l_i \over dx^2} = {\partial V \over \partial l_i},~i=1,2,
\label{deq}
\end{eqnarray}
where $l_i, i=1,2$ are the real and imaginary components of the Polyakov loop.  A domain wall solution ($L(x)$), $L_{ij}$, interpolating $L_i$ and
$L_j$ will satisfy the following
boundary condition,
\begin{eqnarray}
\lim_{x\to-\infty} L(x)=L_i,~\lim_{x\to+\infty} L(x)=L_j.\nonumber
\end{eqnarray}
If we replace the $x-$coordinate by time $t$, then the solution to Eq.\ref{deq} with the above boundary conditions corresponds to the trajectory of a particle under the potential, $-V(L)$, shown in Fig. \ref{fig:epot2}.The trajectory starts out at $L_i$ at $t=-\infty$ and approaches $L_j$ at $t=+\infty$. 

As the particle departs from the point $L_i$, the magnitude ($|L|$) initially decreases, reaching a minimum at the midpoint of the trajectory (i.e., at $t=0$), and then increases again untill it arrives at $L_j$. The mid point represents the core of the domain wall. Along the path, the phase of $L$ undergoes rapid variation near the midpoint. It is evident that the trajectory must be unique as any deviation from it would cause $|L|\rightarrow \infty$ due to the potential, $-V(L)$. For instance, $L(x)\ne 0$ for any $x$. Any trajectory originating from $L_i$ that does pass through $L=0$ would have constant arg($L$), causing the particle to continue in a straight line and inevitably escape to infinity. In standard terminology, this domain wall solution is called a bounce solution. 


In the deconfined phase, let us consider a junction of $L_{01}, L_{12}$ and $L_{20}$ along the $z-$axis. As argued above, except near the $z-$axis, in most of the physical space $|L|$ will take the value corresponding to the minimum of $V(L)$. Though near the domain wall core, $|L|$ will reduce to a finite non-zero value. Thus, the phase of $L$, i.e. $Arg(L)$, is well defined everywhere except near $z-$axis. Across each of the domain wall, $Arg(L)$ varies by $2\pi/3$. So, the three domain walls can be oriented such that the total variation of phase along a hypothetical loop around the $z-$axis  is $\pm2\pi$. 
Note that when the phase of a complex-valued field is well defined along a loop in physical space, its total variation along the loop will always be, $2\pi n$, where $n$ is an integer also known as the winding number. In the current scenario, assuming the Polyakov loop configuration is continuous, the total phase variation remains constant as the loop shrinks toward the $ z$-axis. If $n\ne 0$, below a certain size, the phase variation will lead to an increase in free energy unless $|L|$ decreases. Consequently, as the loop shrinks to a point,  $|L|$ must vanish.  Such a configuration is topologically stable, as small deformations cannot change the winding number. 

\begin{figure}[t]
	\centering
	\includegraphics[width = 0.80\linewidth]{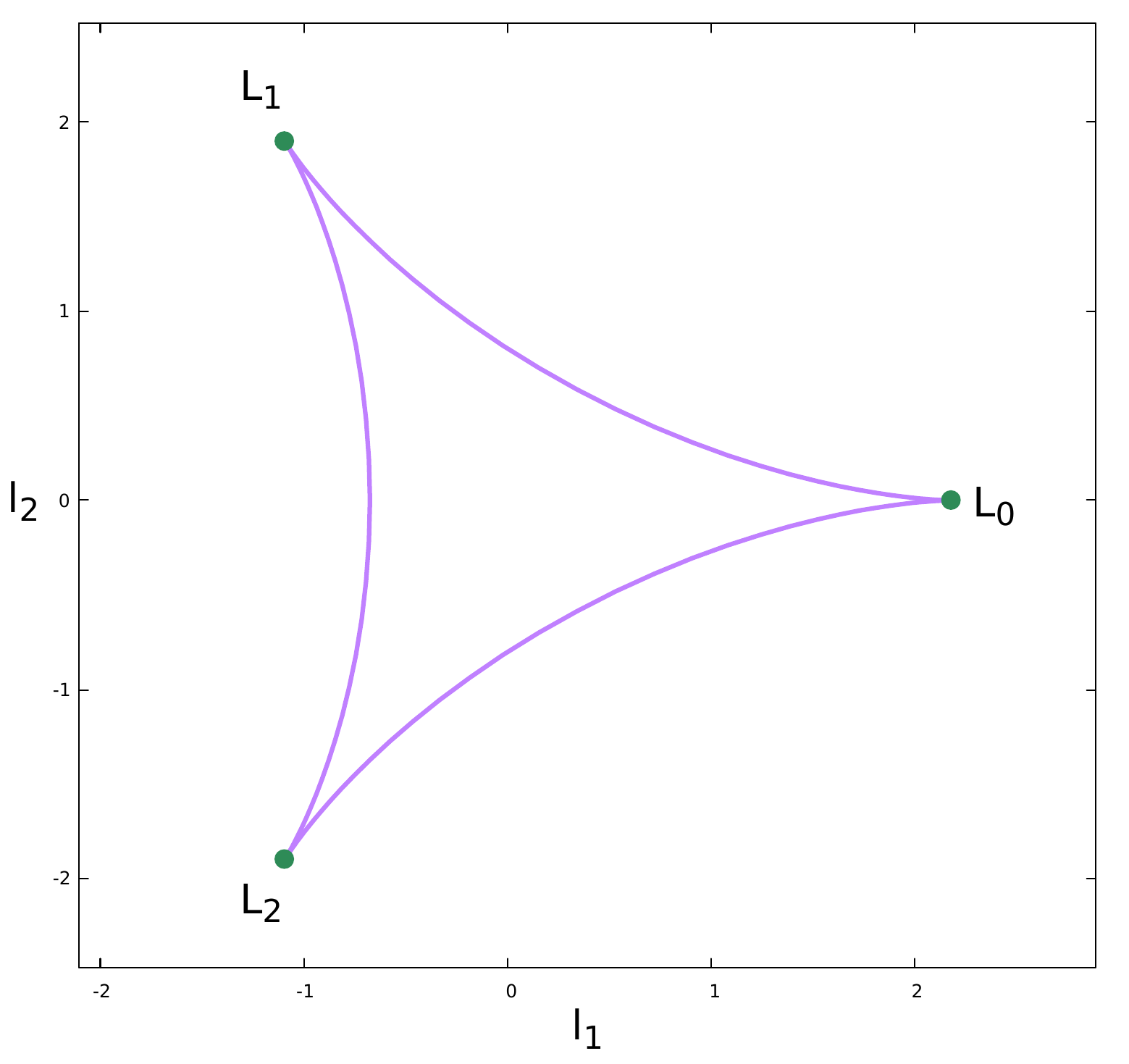}
	\caption{The order parameter space in the presence of domain walls}
	\label{ops}
\end{figure}

Note that conventionally winding numbers, that classify topological strings, are defined when the order parameter space (OPS), consisting of the minima of the effective potential, is a circle ($S^1$). The winding numbers correspond to the mappings from loops in physical space to the OPS, $S^1$, and are classified by the first homotopy group, i.e $\pi_1(S^1)$, which are characterised by set of integers, $\mathbb{Z}$. In the present case, it is useful to define a manifold($M$) consisting of all the possible values of the Polyakov loop, taking into account the presence of domain walls.  According to this definition, $M$ takes the shape of a deformed equilateral triangle as shown in Fig.\ref{ops}. This deformation is due to the profile of the domain walls. Now, one can consider mappings from loops in physical space to $M$.  Since, $M$ is homeomorphic to $S^1$, the corresponding homotopy group will also be $\pi_1(M)=\mathbb{Z}$.
 
As mentioned earlier, the string configuration has been previously studied using the Polyakov loop effective potential. To validate its existence and determine its properties more accurately, it is necessary to simulate the exact partition function with a string in the background, which we undertake in this work. This approach accounts for all possible thermal fluctuations, ensuring more reliable results. In the following, we describe the lattice formulation of the $SU(3)$ gauge theory at finite temperature.

\section{$SU(3)$ gauge theory on the lattice}
We consider the standard Wilson action \cite{Boyd:1996bx}, as the discrete formulation of the action, Eq.\ref{caction} , on a four dimensional Eucledian lattce,
\begin{align}
    S = \beta \sum_{n,\mu>\nu} Tr[1- \frac{1}{3} Re~U_{\mu\nu}],
\label{eq:action}
\end{align}
where $U_{\mu\nu}=U_\mu(n)U_\nu(n+\hat{\mu}){U^{\dagger}}_\mu(n+\hat{\nu}){U^{\dagger}_\nu(n)}$ is the standard Wilson plaquette with $\mu,\nu=1,2,3,4$, $n=(\vec{n},n_4)$. $\hat{\mu}$ is a vector of length equal to the lattice spacing($a$) in the $\mu-$th direction. $U_\mu(n)=\exp[igaA_\mu(n)]$ is the link variable, 
resides on the link connecting the lattice points $n$ and $n+\hat{\mu}$. $\beta=6/g^2$ is the lattice gauge coupling that regulates the lattice
constant ($a$) and the temperature scale through the $\beta-$function.

On the lattice, the trace of the product of links in the temporal direction gives the Polyakov loop operator,
\begin{align}
    L(\vec{n}) = \frac{1}{3} Tr\prod_{n_4=1}^{N_\tau} U_4(\vec{n},n_4)
    \label{3d_pl}
\end{align}
For the realisation of the string configurations, we use the Polyakov loop as a function of two spatial directions (x,y), which is given by,
\begin{align}
    L(n_1,n_2) = \frac{1}{N_z} \sum_{n_3=1}^{N_z} \frac{1}{3} Tr\prod_{n_4=1}^{N_\tau} U_4(n_1,n_2,n_3,n_4),
    \label{2d_pl}
\end{align}
Under gauge transformation, the gauge links  transform as: $U_\mu(n) \rightarrow \Lambda(n)\, U_\mu(n)\, \Lambda^\dagger(n+\hat{\mu})$. As mentioned previously, since the Euclidean time is compactified with a period of $N_\tau$, the gauge transformation is required to be periodic only up to a center element, i.e., $\Lambda(N_\tau,\vec{n}) = z\Lambda(0,\vec{n})$.  Accordingly, the Polayakov loop transforms as $L(n)\rightarrow z L(n)$, where $z\in Z_3$ group {i.e., $z\in 1,\exp(2\pi i/3),\exp(4\pi i/3)$}. At finite temperatures when the gauge fields undergo CD transition, in the deconfined phase, the $Z_3$ symmetry is broken, which is characterised by the non-vanishing value of $L(n)$. The possible values of $z$ allow us to have interfaces and string configurations. 

String configurations are topological structures that are generally unexpected in an equilibrium system due to their associated free energy cost. They form during phase transitions, evolve and subsequently annihilate upon encountering their counterparts. For example, a string and an anti-string (where the total phase variation is $-2\pi$ for a loop traversed clockwise in physical space) will annihilate each other.

To compute the free energy of strings, appropriate boundary conditions are imposed on the lattice to induce a specific string configuration, as discussed in the next section. Specifically, we evaluate the free energy difference between the configurations with and without strings. The free energy can be calculated from the partition function, $F=-T ln Z$, where $Z$ is the partition function and $T$ is the temperature, but in practice, the direct evaluation of the partition function in lattice simulations is avoided as it is difficult. Instead, we evaluate the derivative of the free energy $w.r.t.$ the lattice gauge coupling $\beta$, which is related to the action difference between simulations
with and without the string, i.e.,
\begin{align}
    \frac{\partial}{\partial \beta} \left(\frac{F}{T}\right) &= \frac{1}{\beta} \langle \Delta S \rangle,
\end{align}
where $\Delta S = S_1-S_0$, and $S_1$ and $S_0$ are actions with and without string respectively. The free energy is obtained by integrating the expectation value, 
\begin{eqnarray}
    \frac{F}{T}{\Big|}_{\beta_c}^{\beta}&=& \int_{\beta_c}^{\beta} d\beta~\left <{1 \over \beta}\Delta S\right> \nonumber\\
    &=&\int_{\beta_c}^{\beta} d\beta~\left <\Delta \left(\sum_{n,\mu>\nu} Tr[1- \frac{1}{3} Re~U_{\mu\nu}]\right)\right>
    \label{eq:int_free}
\end{eqnarray}
On a lattice with temporal extension $N_\tau$, the temperature is given by $T=1/(aN_\tau)$.  The string tension, $\sigma$, of the string is defined as free energy per unit length, 
\begin{equation}
 \sigma={F \over L_z}, 
\end{equation}
where $L_z$ is the spatial extension in the z-direction (length of the string). From this, we compute the string tension in units of $T^2$, i.e., $\sigma/T^2$, as a function of $\beta$. Since three or more domain walls emanate from the string and extend to infinity, the string tension will be dependent on the size of the system.

\section{Numerical Setup and Results}

For the simulations, we use lattices of size $N_x=N_y=60$, $N_z=4$ with
$N_\tau=2,4$. The string is aligned along the $z$-direction. We consider
$N_z=4$ as it suppresses fluctuations in the string length. For
$N_\tau=2$ and $4$, the critical couplings for the CD transition are
$\beta_c=5.099$ and $5.6925$, respectively. Since string configurations
exist only in the deconfined phase, we consider $\beta>\beta_c$ in our
simulations.

\begin{table}[htbp]
\centering
\caption{Interface action difference for $N_\tau=2$.}
\label{tab:interface_nt2}
\small
\setlength{\tabcolsep}{6pt}
\begin{tabular}{c cc cc}
\hline\hline
$\beta$ & \multicolumn{2}{c}{patch 1} & \multicolumn{2}{c}{patch 2} \\
\cline{2-3}\cline{4-5}
& $\Delta S_1$ & error & $\Delta S_2$ & error \\
\hline
5.20  & 2.60849 & 0.03857 & 2.55569 & 0.02348 \\
5.30  & 2.05266 & 0.03285 & 2.01550 & 0.02023 \\
5.40  & 1.78793 & 0.03127 & 1.75478 & 0.01932 \\
5.50  & 1.52388 & 0.02907 & 1.51606 & 0.01782 \\
5.60  & 1.32463 & 0.02663 & 1.32239 & 0.01620 \\
5.70  & 1.11286 & 0.02590 & 1.12360 & 0.01620 \\
5.80  & 0.95910 & 0.02502 & 0.96622 & 0.01540 \\
5.90  & 0.86711 & 0.02341 & 0.85801 & 0.01430 \\
6.00  & 0.76401 & 0.02248 & 0.77086 & 0.01371 \\
6.10  & 0.68937 & 0.02193 & 0.69040 & 0.01315 \\
6.20  & 0.64199 & 0.02100 & 0.62903 & 0.01289 \\
6.30  & 0.55285 & 0.02026 & 0.56255 & 0.01237 \\
6.40  & 0.53666 & 0.01950 & 0.54253 & 0.01197 \\
7.00  & 0.43201 & 0.01709 & 0.43442 & 0.01009 \\
8.00  & 0.30046 & 0.01422 & 0.29513 & 0.00844 \\
9.00  & 0.30913 & 0.01227 & 0.31334 & 0.00728 \\
10.00 & 0.28124 & 0.01085 & 0.27210 & 0.00643 \\
11.00 & 0.25258 & 0.00971 & 0.25486 & 0.00575 \\
12.00 & 0.23825 & 0.00889 & 0.24045 & 0.00521 \\
13.00 & 0.23209 & 0.00808 & 0.22244 & 0.00477 \\
14.00 & 0.21455 & 0.00745 & 0.21222 & 0.00445 \\
15.00 & 0.20374 & 0.00696 & 0.19966 & 0.00403 \\
\hline\hline
\end{tabular}
\end{table}

\begin{table}[htbp]
\centering
\caption{Interface action difference for $N_\tau=4$.}
\label{tab:interface_nt4}
\small
\setlength{\tabcolsep}{6pt}
\begin{tabular}{c cc cc}
\hline\hline
$\beta$ & \multicolumn{2}{c}{patch 1} & \multicolumn{2}{c}{patch 2} \\
\cline{2-3}\cline{4-5}
& $\Delta S_1$ & error & $\Delta S_2$ & error \\
\hline
5.80  & 0.22362  & 0.02721 & 0.26723  & 0.01704 \\
5.90  & 0.25909  & 0.02521 & 0.28556  & 0.01566 \\
6.00  & 0.22189  & 0.02370 & 0.23457  & 0.01473 \\
6.10  & 0.22350  & 0.02229 & 0.21348  & 0.01389 \\
6.20  & 0.17070  & 0.02189 & 0.16103  & 0.01359 \\
6.30  & 0.16071  & 0.02124 & 0.15917  & 0.01315 \\
6.40  & 0.14759  & 0.02051 & 0.15209  & 0.01284 \\
7.00  & 0.10863  & 0.01755 & 0.11986  & 0.01088 \\
8.00  & 0.08822  & 0.01492 & 0.08180  & 0.00925 \\
9.00  & 0.08280  & 0.01282 & 0.09236  & 0.00801 \\
10.00 & 0.07491  & 0.01134 & 0.07590  & 0.00701 \\
11.00 & 0.06360  & 0.01006 & 0.07088  & 0.00622 \\
12.00 & 0.05636  & 0.00936 & 0.06180  & 0.00576 \\
\hline\hline
\end{tabular}
\end{table}

For the free energy calculations, we carry out simulations with $\beta$ range
$[5.2, 15.0]$ for $N_\tau=2$ and $[5.8,12.0]$ for $N_\tau=4$. Thermal configurations of link variables are
generated using the Cabibbo-Marinari algorithm
\cite{Cabibbo:1982zn}. Each link variable is updated using this
heat-bath algorithm, which constitutes a single sweep. Since a new
configuration is generated from a previous one, there is always a
non-zero autocorrelation between them. To reduce this correlation, we
perform five heat-bath sweeps between successive measurements. 
For each $\beta$, we sample 5000 configurations for $N_\tau = 2$ and 10000 for 
$N_\tau = 4$ to compute physical observables; the larger sample size for 
$N_\tau = 4$ is required due to stronger fluctuations.

\begin{figure}[htbp]
	\centering
	\includegraphics[width = 0.75\linewidth]{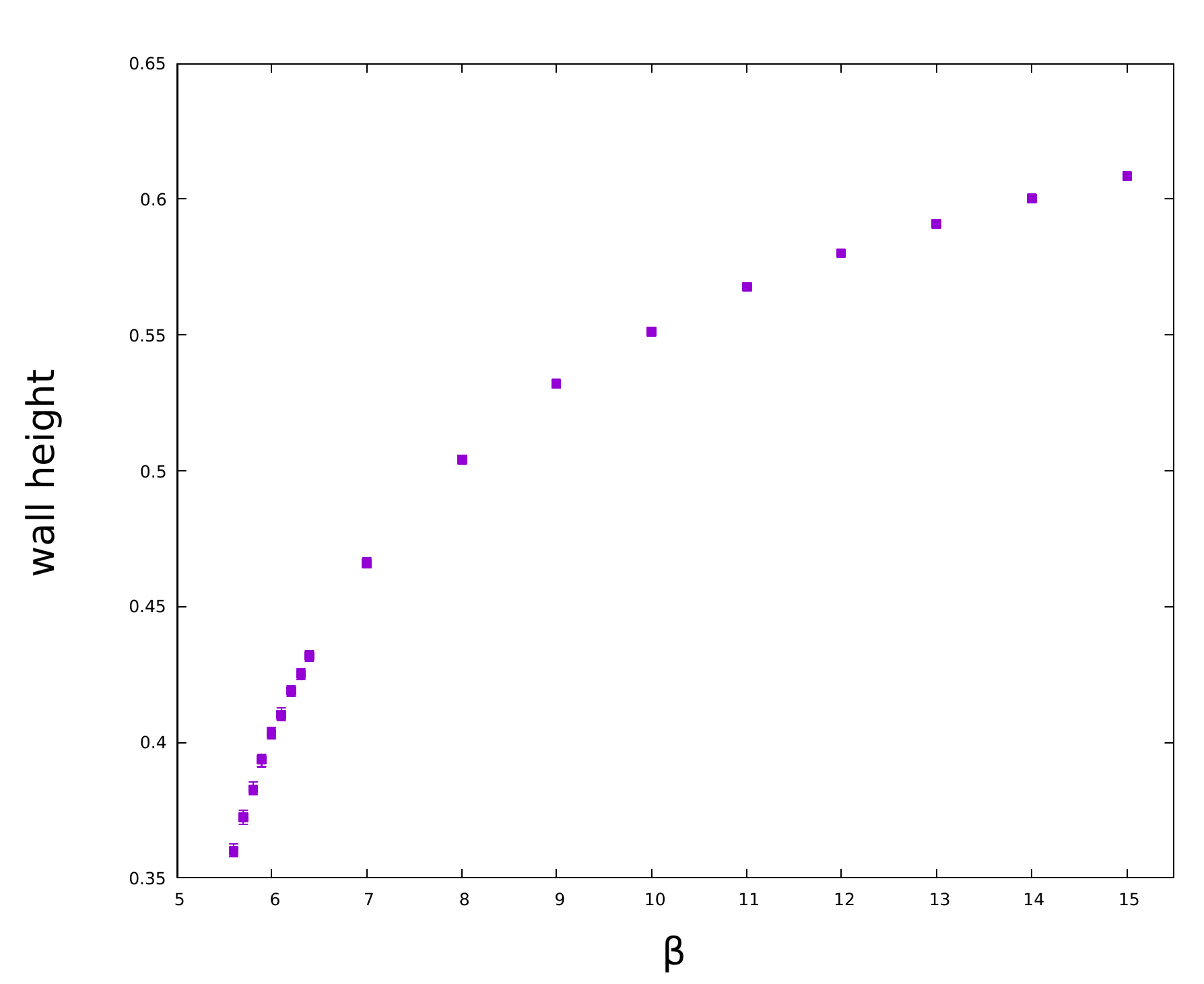}
	\caption{The interface wall height $vs$ $\beta$ for $N_\tau=2$}
	\label{fig:wallheight_beta}
\end{figure}

As discussed earlier, in effective models, in the presence of domain
walls the manifold $M$ takes the form of a deformed triangle
(Fig.\ref{ops}). To establish that the string configuration remains
topological in the exact theory, it is necessary to show that $M$ is a
deformed triangle of non-zero size in the complex Polyakov loop plane.
The size of $M$ is determined by the bulk equilibrium value of the
Polyakov loop (largest) and the value at the core of the domain walls
(smallest). As long as the smallest value remains non-zero, the size of
$M$ is non-zero and the corresponding string configurations are
topological.

\begin{figure}[h]
	\centering
	\includegraphics[width = 0.90\linewidth]{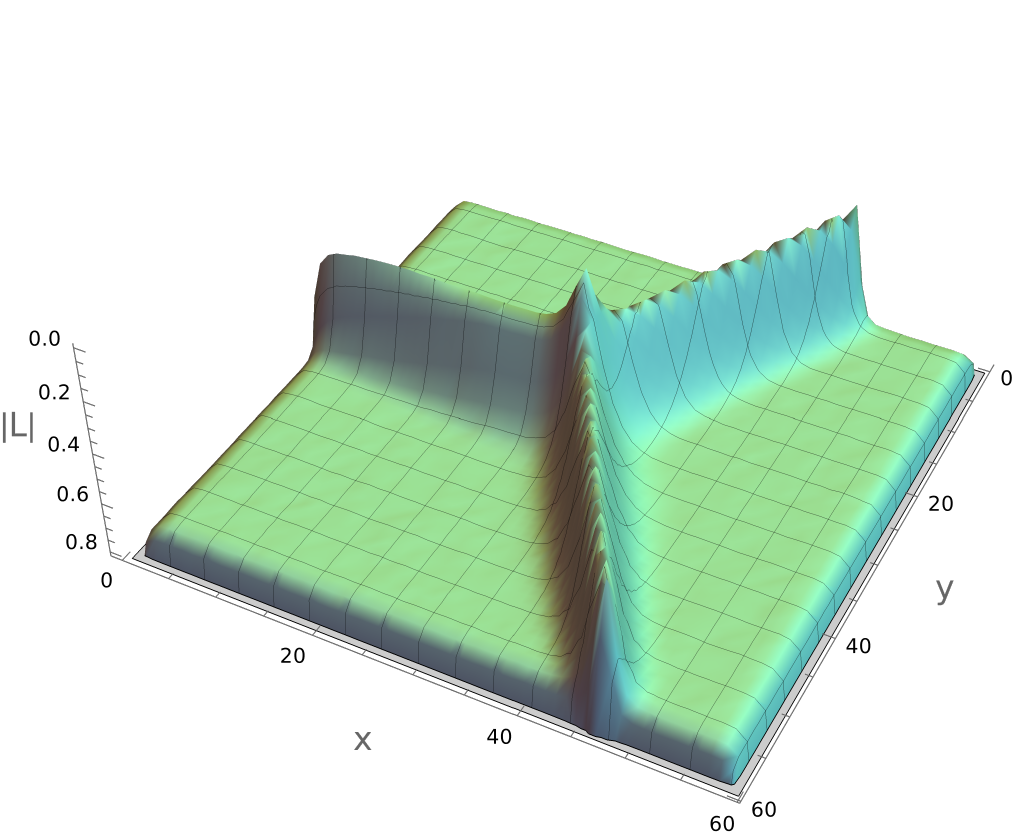}
	\includegraphics[width = 0.70\linewidth]{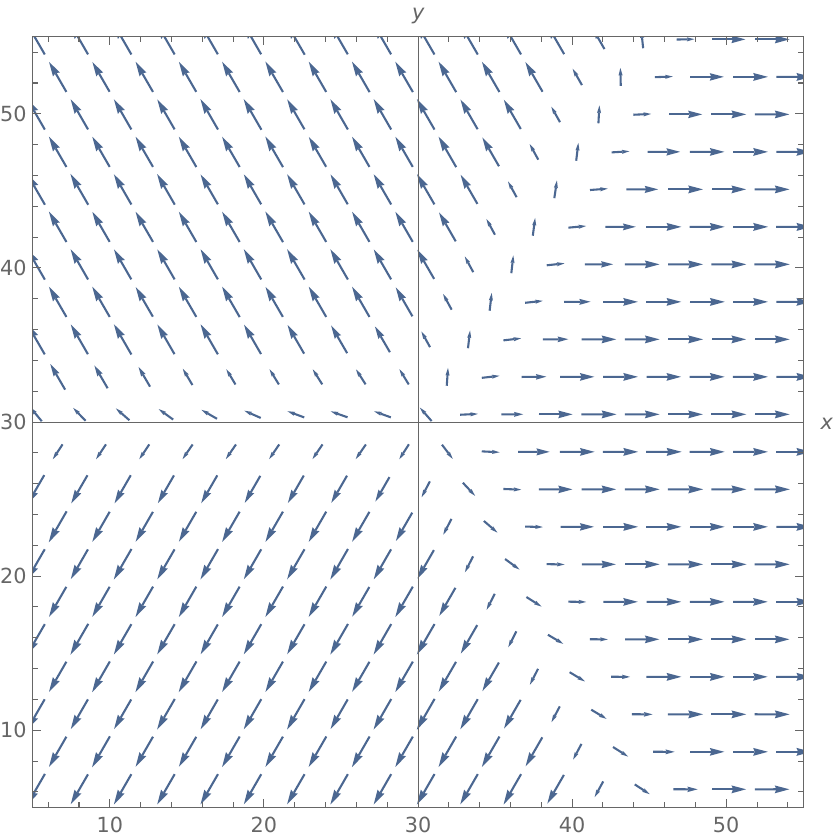}
	\caption{A typical string attached to $Z_3$ domain walls for $\beta=7.0$ for $N_\tau=2$. The absolute value of the Polyakov loop as a function of (x,y)(top), A vector plot with the real and imaginary part of the Polyakov loop(bottom).}
	\label{fig:string_config_nt2}
\end{figure}

\begin{figure}[h]
	\centering
	\includegraphics[width = 0.90\linewidth]{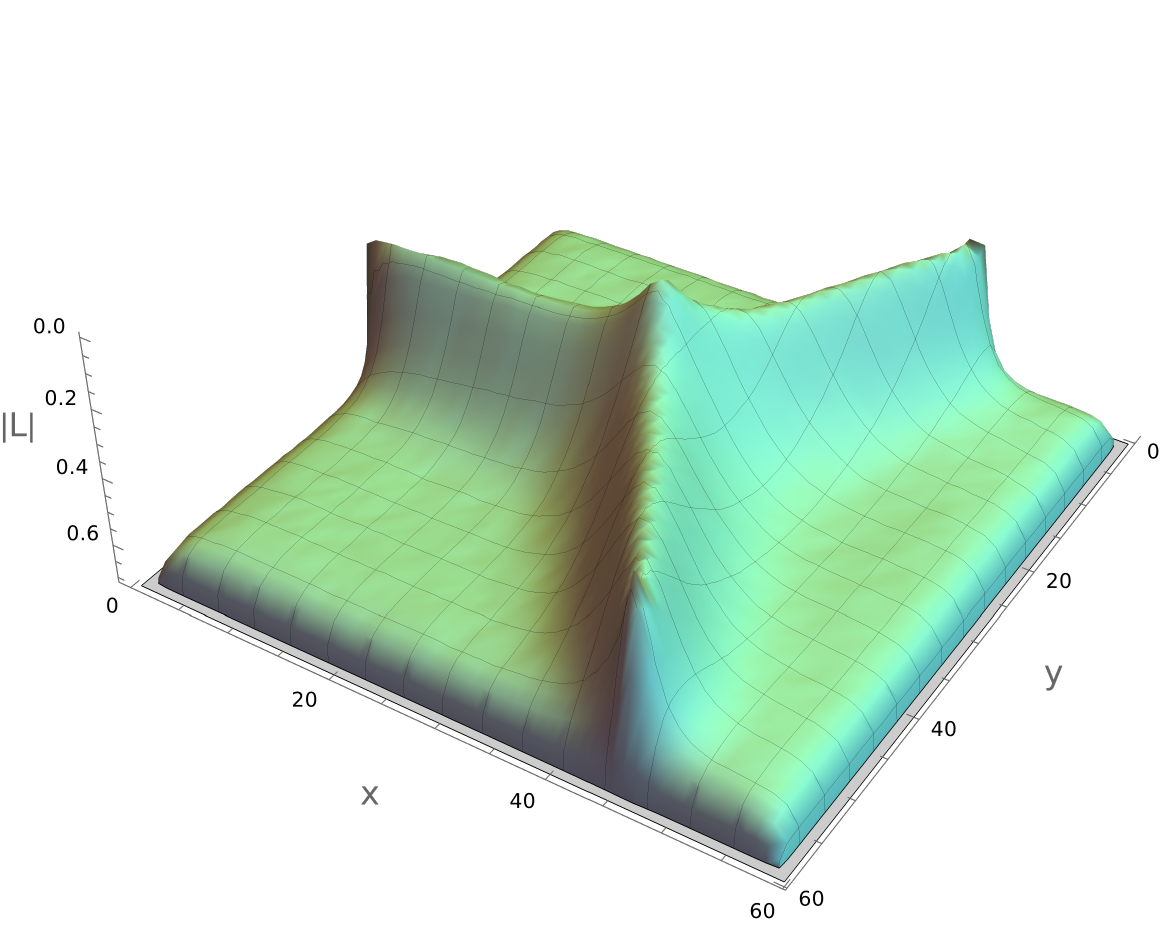}
	\includegraphics[width = 0.70\linewidth]{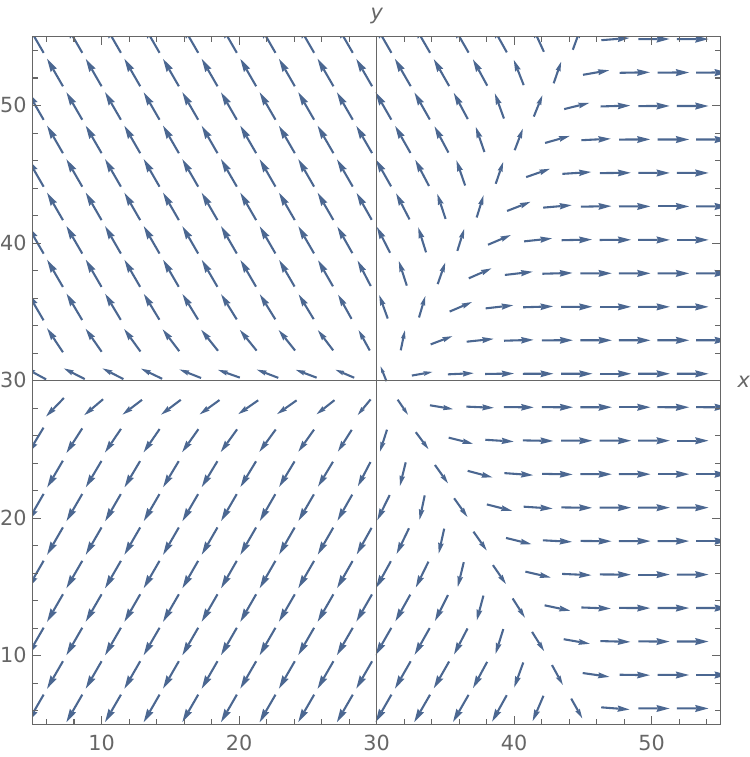}
	\caption{A typical string attached to $Z_3$ domain walls for $\beta=10.0$ for $N_\tau=4$. The absolute value of the Polyakov loop as a function of (x,y)(top), A vector plot with the real and imaginary part of the Polyakov loop(bottom).}
	\label{fig:string_config_nt4}
\end{figure}

We therefore calculate the magnitude of the Polyakov loop at the core
of the domain walls. The results for the Polyakov loop value at the
center of the domain wall for different $\beta$ are shown for $N_\tau=2$ in
Fig.\ref{fig:wallheight_beta}. We mention here that the Polyakov loop magnitude, at the core 
of the domain walls, decreases
with decrease in temperature faster than the thermal average of the Polyakov loop. 
This faster decrease especially near $T_c$ could be due to perfect wetting, 
i.e nucleation of confinement-deconfinement interfaces.
The results of Fig.\ref{fig:wallheight_beta} show that the Polyakov loop
at the core of domain walls is indeed non-zero. Thus the $Z_3$ strings
resulting from the junctions of these walls are topological, as they
correspond to non-trivial mappings from loops in physical space to
$M$. Consequently, we observe that the Polyakov loop vanishes at the
core of the $Z_3$ strings, see Fig.\ref{fig:string_config_nt2} and 
Fig.\ref{fig:string_config_nt4}.

The calculation of the free energy of the string configuration requires
simulations both with and without the string. This involves determining
the difference in action between the two cases, $\Delta S$. In the
simulations with the string, the temporal links are set according to
$n_4$. For $n_4 < N_\tau$, all the temporal links are set to
$U_4(n)=\mathds{1}$. For $n_4= N_\tau$, they are set according to the
azimuthal angle coordinate ($\theta_a$) corresponding to the position
vector $\vec{n}$. $U_4(n)=\mathds{1}$ for $-\pi/3\le\theta_a\le\pi/3$
and $U_4(n)=z^{-1}(z)$ for
$-\pi(\pi/3)<\theta_a<-\pi/3(\pi)$, where $z=\exp(2\pi i/3)$. 

\begin{figure}[h]
\centering
\includegraphics[width = 0.75\linewidth]{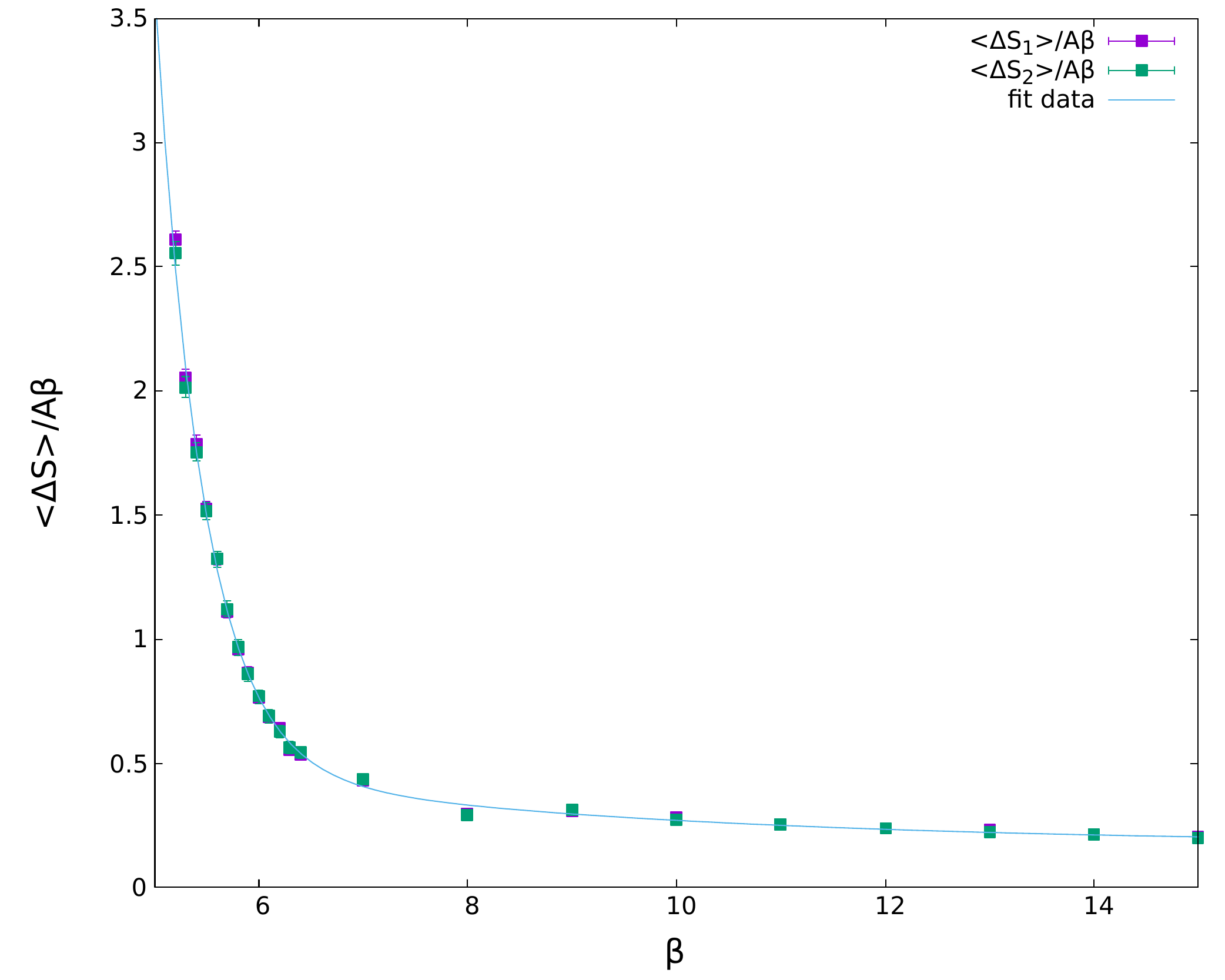}
\caption{$\Delta S / A\beta$ vs. $\beta$ for the $Z_3$ interface with fitted curve, $N_\tau=2$}
\label{fig:dS_interface_beta_nt2}
\end{figure}

\begin{figure}[h]
\centering
\includegraphics[width = 0.75\linewidth]{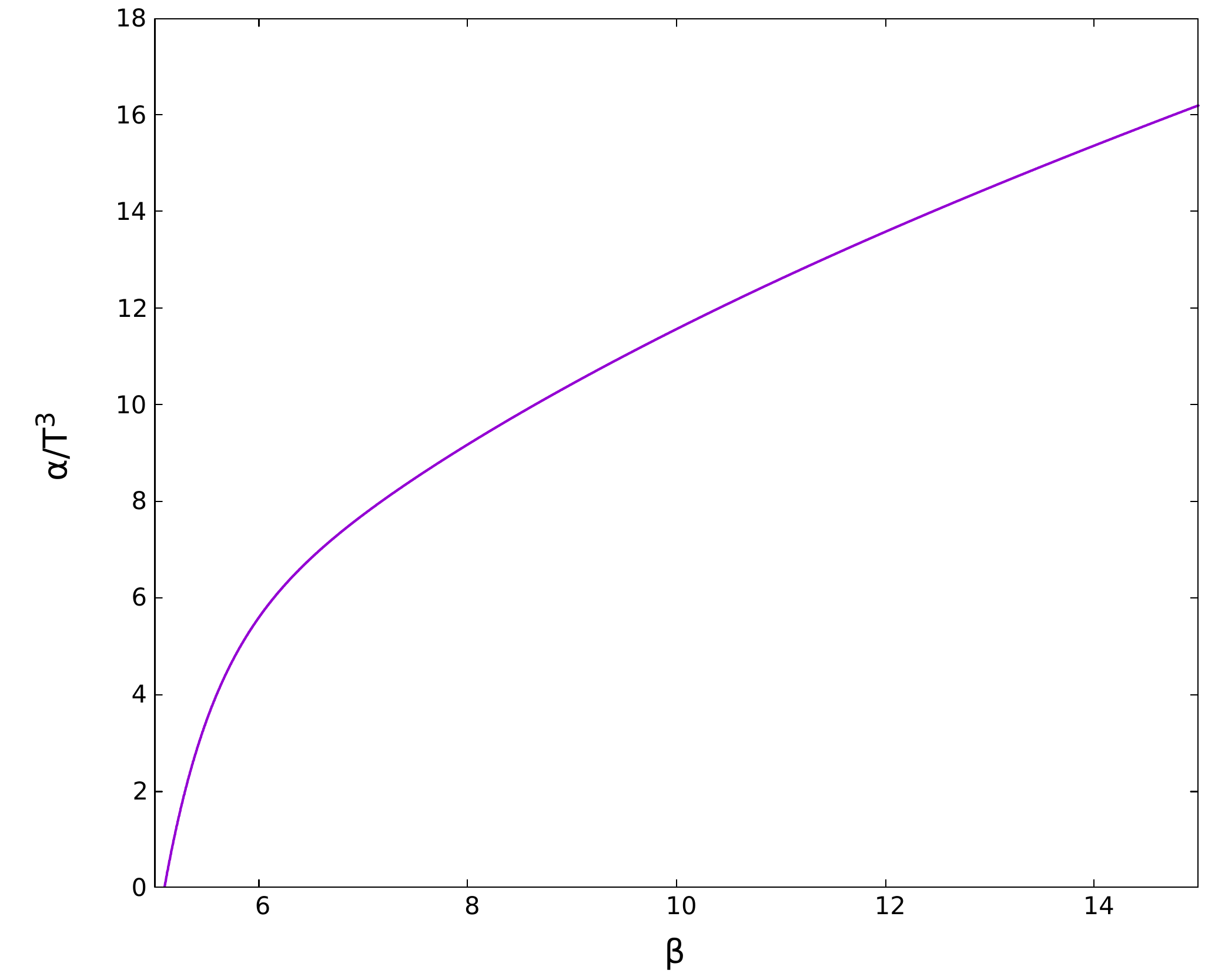}
\caption{$\beta$ dependence of $\alpha/T^3$ plot for $N_\tau=2$}
\label{fig:alphaT_beta_nt2}
\end{figure}

\begin{figure}[h]
\centering
\includegraphics[width = 0.75\linewidth]{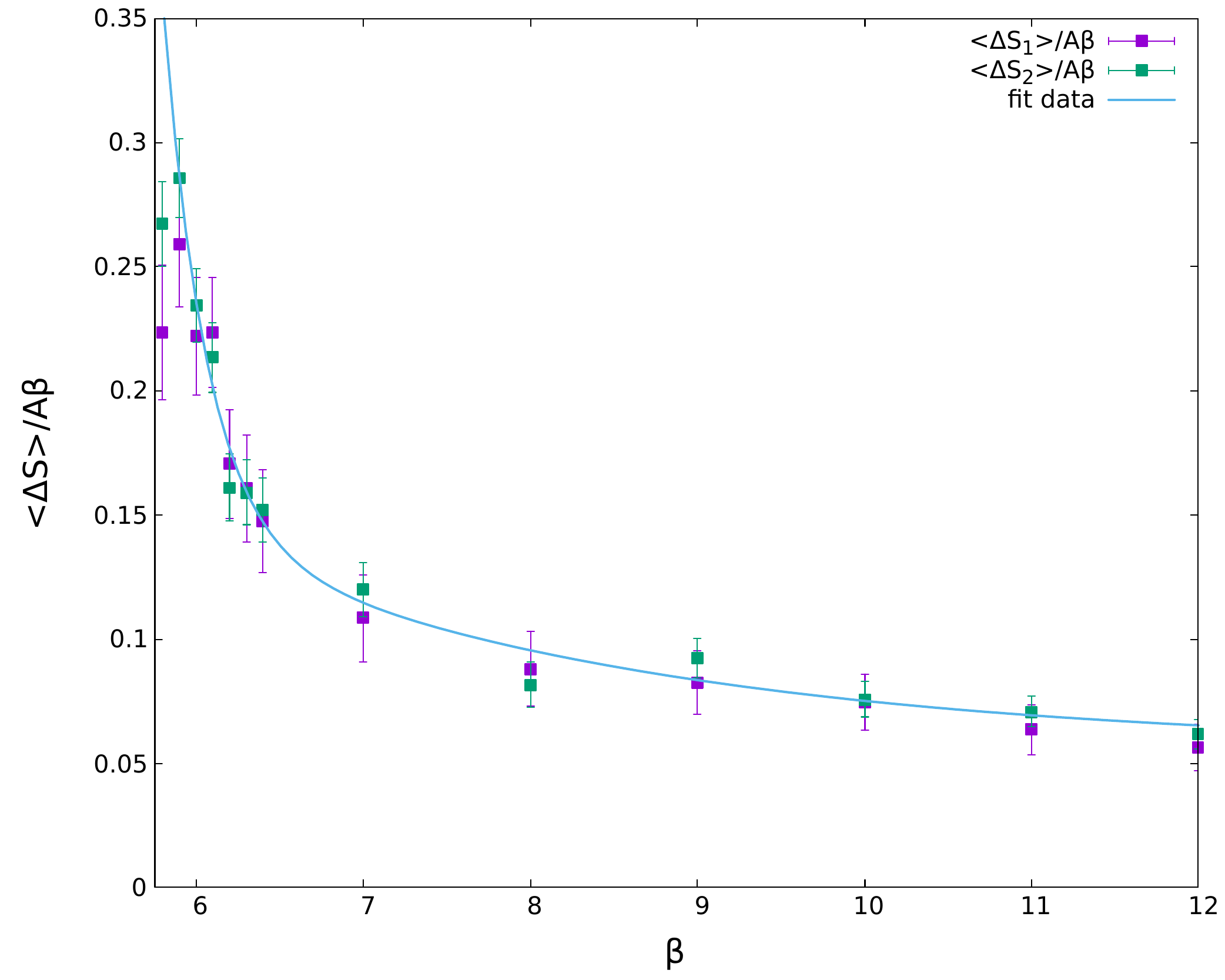}
\caption{$\Delta S / A\beta$ vs. $\beta$ for the $Z_3$ interface with fitted curve, $N_\tau=4$}
\label{fig:dS_interface_beta_nt4}
\end{figure}

\begin{figure}[h]
\centering
\includegraphics[width = 0.75\linewidth]{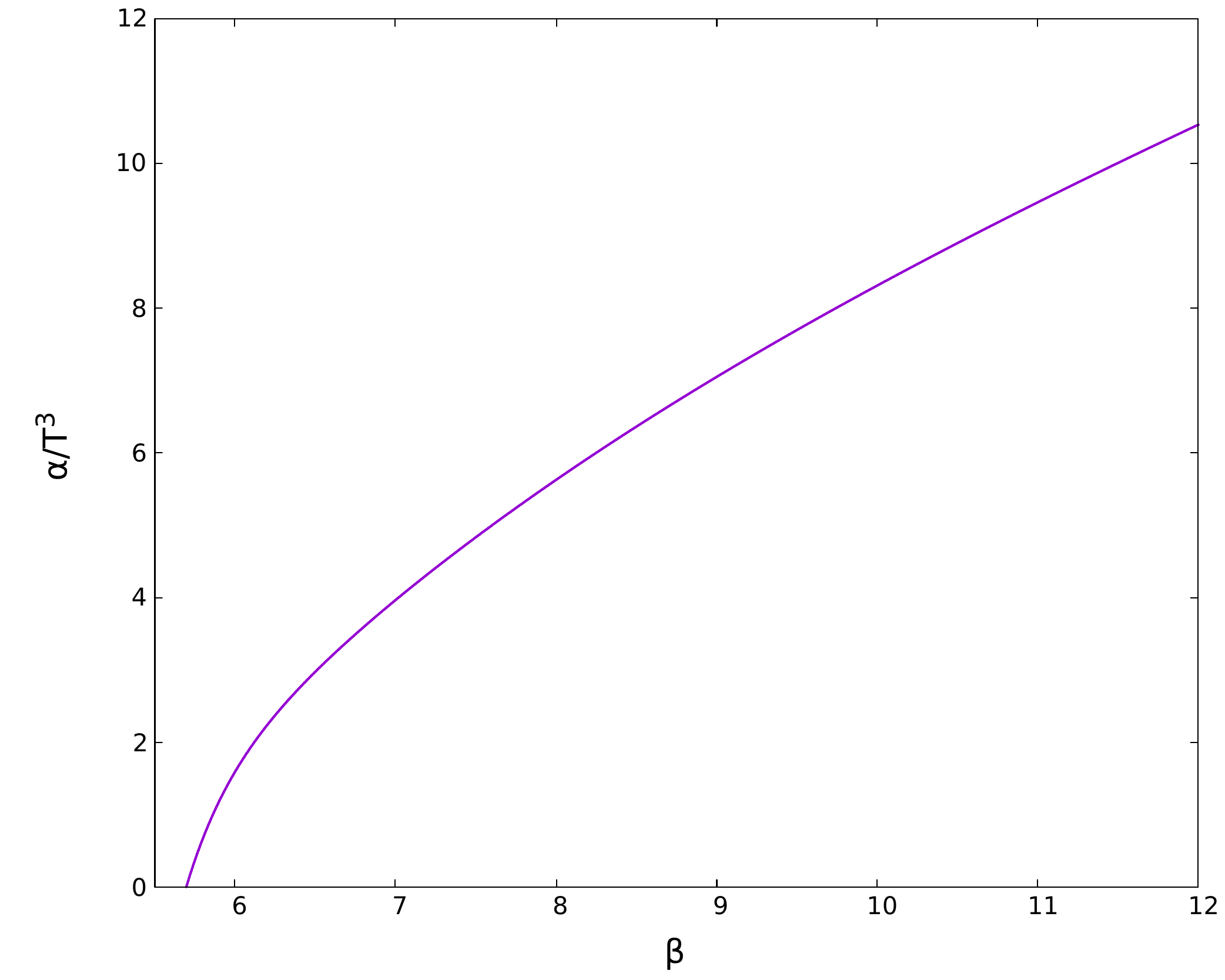}
\caption{$\beta$ dependence of $\alpha/T^3$ plot for $N_\tau=4$}
\label{fig:alphaT_beta_nt4}
\end{figure}

Keeping the temporal links at the boundary fixed ensures that the
initial configuration thermalizes into a string with domain walls
attached. Further, the boundary conditions ensures that 
the average angular separation between the domain walls is $2\pi/3$, 
corresponding to the lowest free energy configurations.
A typical profile of the string at $\beta=7.0$ for $N_\tau=2$ and at 
$\beta=10.0$ for $N_\tau=4$ is shown in Fig.\ref{fig:string_config_nt2} and 
Fig.\ref{fig:string_config_nt4} respectively, where the Polyakov loop is averaged over
the $z$-direction. The string configuration shows that the domain walls broaden near the 
string junction. Our observations indicate that this is due to presence of other domain walls at 
the junction. This broadening resembles the broadening observed close to the 
confinement-deconfinement transition due to perfect wetting. 
The effect of this boundary condition on the magnitude profile of the
 Polyakov loop, decays within a few lattice points from the
boundary wall. To eliminate boundary effects, we exclude the region
near the boundary where physical observables deviate from their bulk
equilibrium values.

In $SU(N)$ gauge theories with $N>2$, when the $Z_N$ symmetry is
spontaneously broken, the string configuration is always attached to
$N$ domain walls. The interface tension of these walls contributes
significantly to the total string tension. Therefore before discussing
the string configuration, we first present our results for the interface
tension of the domain walls connected to the string configurations.

We calculate the interface tension away from the string core. This is
because the magnitude of the Polyakov loop vanishes at the core
(Fig.\ref{fig:string_config_nt2} and Fig.\ref{fig:string_config_nt4}), causing the 
domain walls to deform in that region. To avoid these effects we consider 
the action difference
$\Delta S$ within radial annular regions around the string core,
defined by $r_a\le r/a\le r_b$, with $r_a=10$ and $r_b=15,20$. 
We then compute the action difference, divide it by three (corresponding to the
three interfaces), and normalize by the area of the annular patches as
well as the area of the domain walls. The data for the interface action difference are presented in Tables~\ref{tab:interface_nt2} and \ref{tab:interface_nt4}. The corresponding plots as a function of $\beta$ are shown in Figs.~\ref{fig:dS_interface_beta_nt2} and \ref{fig:dS_interface_beta_nt4} for $N_\tau=2$ and $4$, respectively. In the tables, the results for the patch with $r_b=15$($r_b=20$) are referred to as patch~1(patch~2).
In the plot, $\Delta S_1$ and $\Delta S_2$ 
correspond to $r_b=20$ and $15$ respectively. 
The weak coupling analysis suggests that for large $\beta$ the action difference is expected 
to behave as $\sim 1/\sqrt{\beta}$, \cite{Bhattacharya:1992qb}. We find that our measured values of action differences, are 
compatible with the weak coupling results. However, to fit the data in the whole range of 
$\beta$, we consider the following function
$f(\beta)=b_1\exp[-c_1(\beta-d_1)]
        +b_2\exp[-c_2(\beta-d_2)] + e_1$.
We then integrate this fitted function to calculate the free energy of the interface 
configuration.

For $\beta$ close to $\beta_c$ there are large fluctuations in the
location of the string core, leading to larger errors in the action
difference. Nevertheless, we obtain a reasonable fit to the data very
close to $\beta_c$. The results for the interface tension $\alpha/T^3$
as a function of $\beta$ are shown in Fig.\ref{fig:alphaT_beta_nt2} and 
Fig.\ref{fig:alphaT_beta_nt4} corresponding to $N_\tau=2$ and $4$. The 
results for $N_\tau=2$ qualitatively agree with previous 
work \cite{Kajantie:1990bu}, 
confirming the reliability of our method.

It is well known that the pure $SU(3)$ CD transition is weakly
first-order. At $\beta=\beta_c$ the confined state ($L=0$) and the
three ($L\neq0$) $Z_3$ states are degenerate. As a result, in addition
to $Z_3$ domain walls, CD interfaces also emerge. Effective model
calculations indicate that near the transition point these domain walls
decay into pairs of CD interfaces. This behaviour is also observed in
the lattice simulations, where due to {\it perfect wetting} locally
CD interfaces can nucleate near the domain wall core, effectively 
broadening the domain walls and eventually, for temperatures near $T_c$, 
lead to the decay of domain walls into CD interfaces, 
\cite{deForcrand:2004jt,Frei:1989es}. Thus at $\beta=\beta_c$ 
the interface tension is expected to be 
approximately twice that of the CD interface.
In the following we present
our results for the string tension for $N_\tau=2$ and $N_\tau=4$.

\subsection{$N_\tau=2$ results}

\begin{figure}[h]
\centering
\includegraphics[width = 0.75\linewidth]{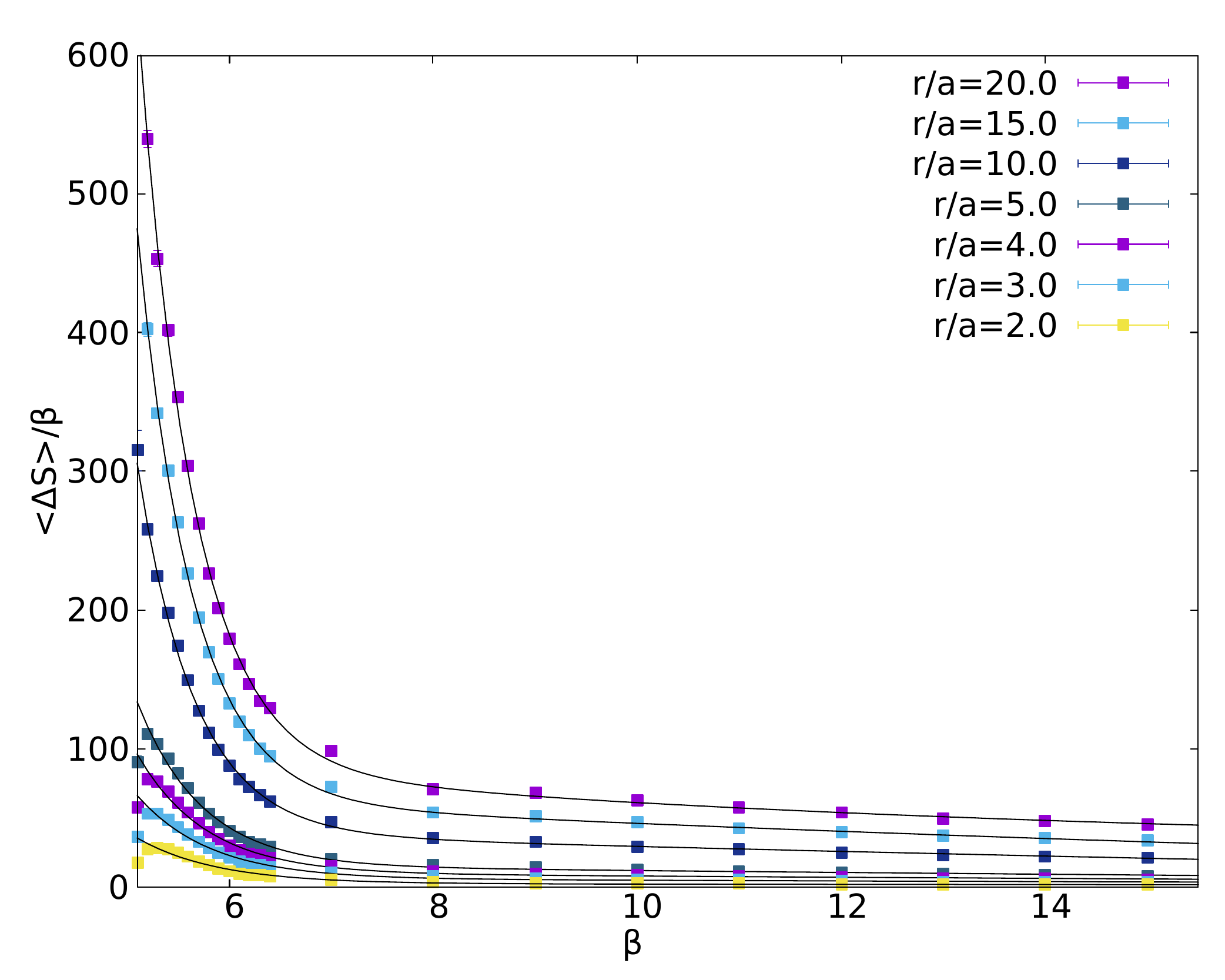}
\caption{ For $N_\tau=2$, $\Delta S / \beta$ vs. $\beta$ for different radial patches 
$r/a=5,10,15,20$ fitted curve with function $f(x)$.}
\label{fig:dS_beta_nt2}
\end{figure}

For $N_\tau=2$, we compute the action difference $\Delta S$ corresponding
to different radial patches around the string core. The results for
$\Delta S/\beta$ as a function of $\beta$ are shown in
Fig.\ref{fig:dS_beta_nt2} and the corresponding data is given in Table~\ref{tab:string_action_nt2}. 
We mention here that the action difference will constitute
the contribution from the domain walls as well as the string core. Unlike the contribution 
from the domain walls, there are no known results for the string core in the
weak coupling limit.
However, reasonable fit to the data for the whole range of $\beta$ considered for simulations
was achieved with the function  
$f(\beta)$ defined earlier. For smaller patches, as $\beta$ approaches
$\beta_c$ from above, the rise in $\Delta S$ stops and decreases to a
smaller non-zero value at $\beta_c$. 
However, this change occurs in a
very narrow region close to $\beta_c$ and does not affect the results
qualitatively.

\begin{figure}[h]
\centering
\includegraphics[width = 0.75\linewidth]{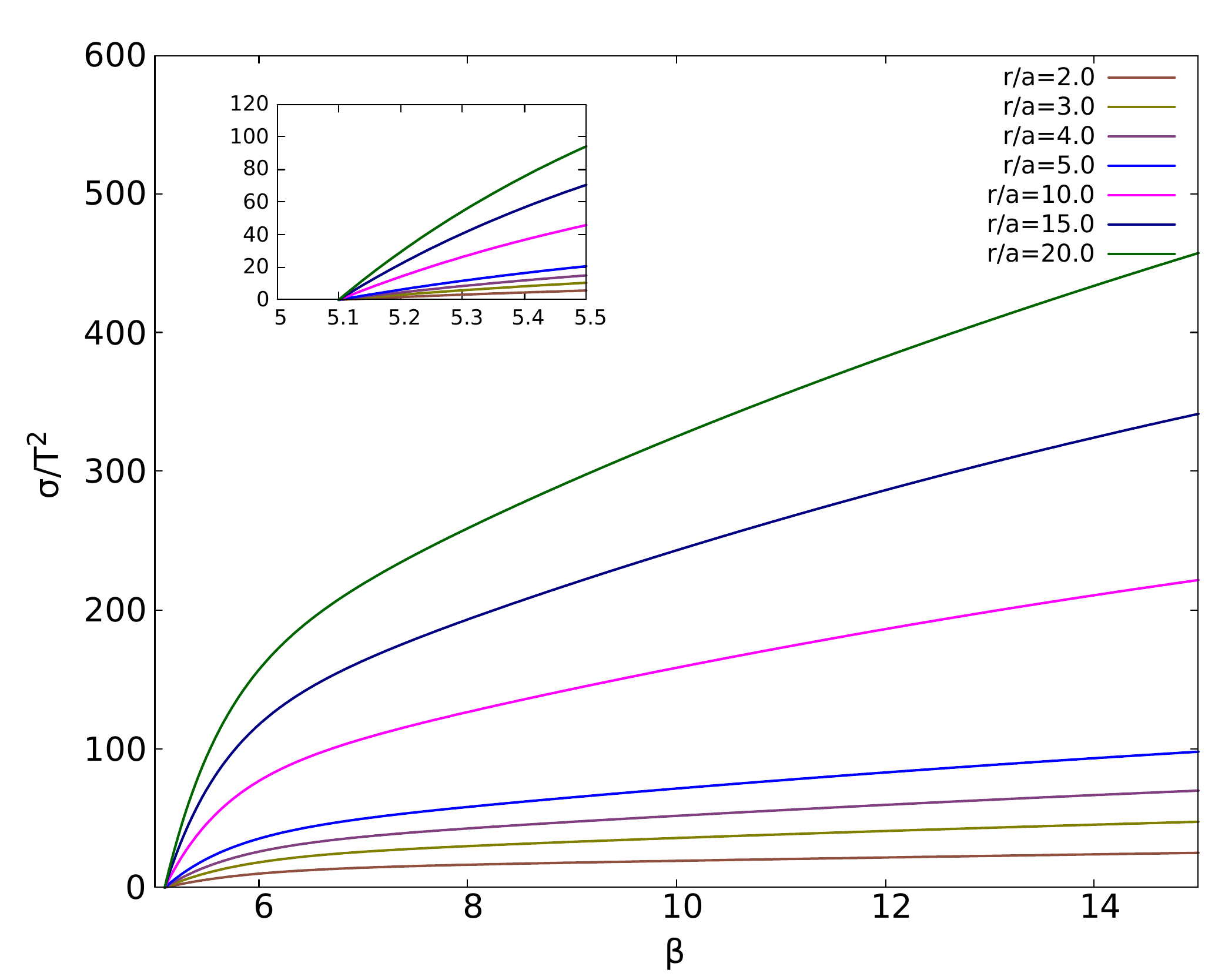}
\caption{For $N_\tau=2$, $\beta$ dependence of $\sigma/T^2$ plot for different radial patches $r/a=5,10,15,20$.}
\label{fig:sigmaT_beta_nt2}
\end{figure}

Following this, we integrate $f(\beta)$ using
Eq.\ref{eq:int_free} to obtain the free energy of the string
configuration. The resulting string tension $\sigma/T^2$ as a function
of $\beta$ for different radial patches is shown in
Fig.\ref{fig:sigmaT_beta_nt2}. A steep rise in $\sigma/T^2$ is observed
close to $\beta_c$, followed by an approximately linear increase for
larger $\beta$.

\begin{center}
\begin{table*}[htbp]
\centering
\caption{Action difference $\Delta S$ and jackknife error for $N_\tau=2$}
\label{tab:string_action_nt2}
\small
\setlength{\tabcolsep}{5pt}
\begin{tabular}{c cc cc cc cc}
\hline\hline
$\beta$
& \multicolumn{2}{c}{$r/a=5$}
& \multicolumn{2}{c}{$r/a=10$}
& \multicolumn{2}{c}{$r/a=15$}
& \multicolumn{2}{c}{$r/a=20$} \\
\cline{2-3}\cline{4-5}\cline{6-7}\cline{8-9}
& $\Delta S$ & $error$
& $\Delta S$ & $error$
& $\Delta S$ & $error$
& $\Delta S$ & $error$ \\
\hline
5.2  & 110.36484 & 0.60116 & 257.87405 & 1.25330 & 402.49983 & 1.86822 &  539.88657 & 0.41658 \\
5.3  & 103.59388 & 0.54066 & 224.67773 & 1.07956 & 341.85150 & 1.61047 &  453.54355 & 0.38919 \\
5.4  &  92.62027 & 0.50698 & 197.82285 & 1.00640 & 300.20597 & 1.50576 &  401.59460 & 0.32945 \\
5.5  &  82.39651 & 0.47625 & 174.47203 & 0.95316 & 263.79369 & 1.39979 &  353.10258 & 0.30470 \\
5.6  &  71.70216 & 0.45873 & 149.49289 & 0.90595 & 226.07948 & 1.33539 &  303.65829 & 0.28723 \\
5.7  &  61.30492 & 0.43776 & 127.88278 & 0.86314 & 194.87043 & 1.26551 &  262.55875 & 0.27555 \\
5.8  &  53.40427 & 0.40917 & 111.89299 & 0.81756 & 169.56406 & 1.20557 &  226.53799 & 0.26649 \\
5.9  &  47.13730 & 0.40463 &  99.47629 & 0.79970 & 150.55162 & 1.15040 &  201.96085 & 0.25441 \\
6.0  &  41.00473 & 0.37837 &  88.08466 & 0.76249 & 132.83176 & 1.12734 &  179.68027 & 0.24403 \\
6.1  &  36.47997 & 0.37304 &  78.15178 & 0.72854 & 119.75076 & 1.10030 &  161.14301 & 0.23580 \\
6.2  &  33.01000 & 0.36098 &  72.43353 & 0.70372 & 109.79351 & 1.05420 &  147.18480 & 0.22791 \\
6.3  &  31.43271 & 0.34900 &  66.88779 & 0.69024 & 100.48857 & 1.02028 &  134.15992 & 0.22342 \\
6.4  &  29.26770 & 0.34022 &  62.58853 & 0.66935 &  94.97354 & 0.98328 &  128.83123 & 0.21129 \\
7.0  &  20.97773 & 0.28890 &  47.49854 & 0.58267 &  72.86676 & 0.84560 &   98.35134 & 0.18141 \\
8.0  &  16.62055 & 0.24176 &  35.60965 & 0.48075 &  53.88290 & 0.70482 &   71.32716 & 0.15119 \\
9.0  &  14.75837 & 0.21165 &  33.48775 & 0.41564 &  51.24049 & 0.60717 &   68.67947 & 0.12853 \\
10.0 &  13.04334 & 0.18806 &  29.76977 & 0.37084 &  46.83962 & 0.53784 &   62.64910 & 0.11453 \\
11.0 &  11.74472 & 0.17133 &  27.68351 & 0.33805 &  42.57028 & 0.49007 &   57.46369 & 0.10168 \\
12.0 &  10.89854 & 0.15542 &  25.42321 & 0.29672 &  39.76455 & 0.43152 &   54.15185 & 0.09198 \\
13.0 &   9.85393 & 0.14322 &  23.71855 & 0.27401 &  37.22653 & 0.39858 &   50.06809 & 0.08413 \\
14.0 &   9.29583 & 0.13142 &  22.70685 & 0.25704 &  35.26673 & 0.37050 &   48.20444 & 0.07886 \\
15.0 &   8.54539 & 0.11975 &  21.33430 & 0.23581 &  33.56463 & 0.34129 &   45.18588 & 0.07070 \\
\hline\hline
\end{tabular}
\end{table*}
\end{center}

\begin{center}
\begin{table*}[htbp]
\centering
\caption{Action difference $\Delta S$ and jackknife error for $N_\tau=4$}
\label{tab:string_action_nt4}
\small
\setlength{\tabcolsep}{5pt}
\begin{tabular}{c cc cc cc cc}
\hline\hline
$\beta$
& \multicolumn{2}{c}{$r/a=5$}
& \multicolumn{2}{c}{$r/a=10$}
& \multicolumn{2}{c}{$r/a=15$}
& \multicolumn{2}{c}{$r/a=20$} \\
\cline{2-3}\cline{4-5}\cline{6-7}\cline{8-9}
& $\Delta S$ & $error$
& $\Delta S$ & $error$
& $\Delta S$ & $error$
& $\Delta S$ & $error$ \\
\hline
5.8  &  5.68990 & 1.19789 & 15.59788 & 2.50520 & 29.01490 & 4.01806 & 47.66482 & 5.53609 \\
5.9  &  7.31175 & 0.89521 & 19.02333 & 1.65565 & 34.56860 & 2.45648 & 53.29115 & 3.51277 \\
6.0  &  5.91528 & 0.76407 & 16.92231 & 1.54608 & 30.23584 & 2.45368 & 45.07127 & 3.11033 \\
6.1  &  5.58601 & 0.72473 & 14.89238 & 1.40068 & 28.30254 & 2.10851 & 40.51041 & 2.61986 \\
6.2  &  6.07489 & 0.64556 & 15.42404 & 1.29691 & 25.66628 & 1.98814 & 34.74756 & 2.53741 \\
6.3  &  5.56974 & 0.57049 & 15.55875 & 1.17513 & 25.20144 & 1.86186 & 34.65953 & 2.21769 \\
6.4  &  4.78848 & 0.52888 & 13.67205 & 1.16905 & 22.52737 & 1.81993 & 31.92254 & 2.50208 \\
7.0  &  4.38683 & 0.55541 & 10.79059 & 1.03709 & 17.30854 & 1.51373 & 25.17335 & 1.92691 \\
8.0  &  2.91041 & 0.43509 &  9.39564 & 0.83151 & 14.68870 & 1.19849 & 19.21144 & 1.65528 \\
9.0  &  2.28381 & 0.36117 &  7.60547 & 0.72997 & 12.57344 & 1.13864 & 18.68845 & 1.59952 \\
10.0 &  2.39689 & 0.33818 &  6.36685 & 0.64099 & 10.86126 & 1.05836 & 15.47505 & 1.37010 \\
11.0 &  2.02822 & 0.28034 &  6.00958 & 0.54015 &  9.82546 & 0.84510 & 14.51481 & 1.09442 \\
12.0 &  1.59492 & 0.24851 &  4.90090 & 0.48042 &  8.28244 & 0.75817 & 12.31742 & 1.04125 \\
\hline\hline
\end{tabular}
\end{table*}
\end{center}

\subsection{$N_\tau=4$ results}

\begin{figure}[htbp]
\centering
\includegraphics[width = 0.75\linewidth]{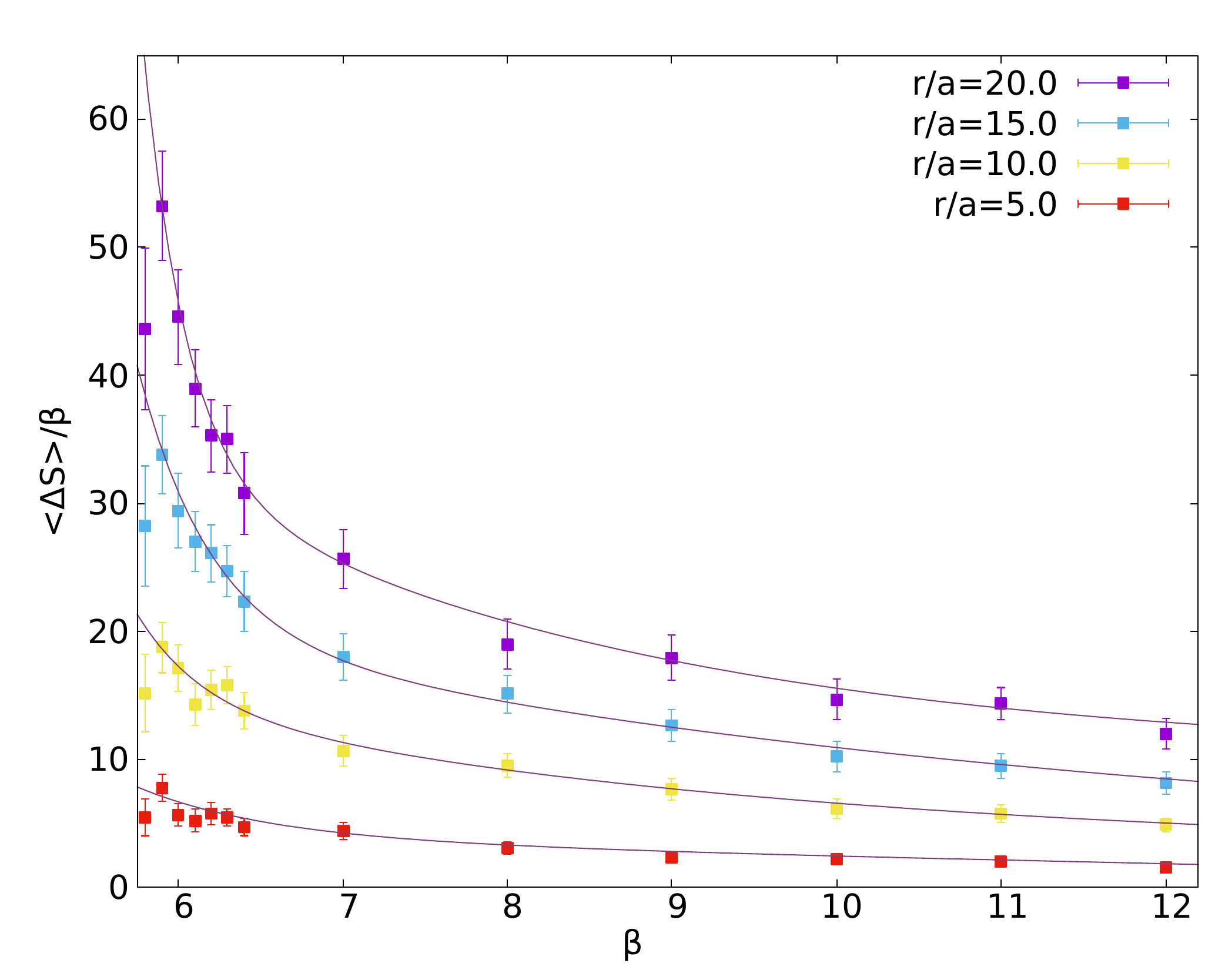}
\caption{For $N_\tau=4$, $\Delta S / \beta$ vs. $\beta$ for different radial patches $r/a=5,10,15,20$ fitted curve with function $f(x)$.}
\label{fig:dS_beta_nt4}
\end{figure}

We repeat the same analysis for $N_\tau=4$ lattices. The action
difference $\Delta S$ is calculated for the radial patches around
the string core Fig.\ref{fig:dS_beta_nt4} and fitted using the same function $f(\beta)$ and the corresponding data is given in Table~\ref{tab:string_action_nt4}. The
qualitative behaviour of $\Delta S$ as a function of $\beta$ remains
similar to that observed for $N_\tau=2$. Near $\beta_c$ the fluctuations
in the string core position again lead to larger statistical errors,
while for larger $\beta$ the behaviour becomes smoother.

\begin{figure}[htbp]
\centering
\includegraphics[width = 0.75\linewidth]{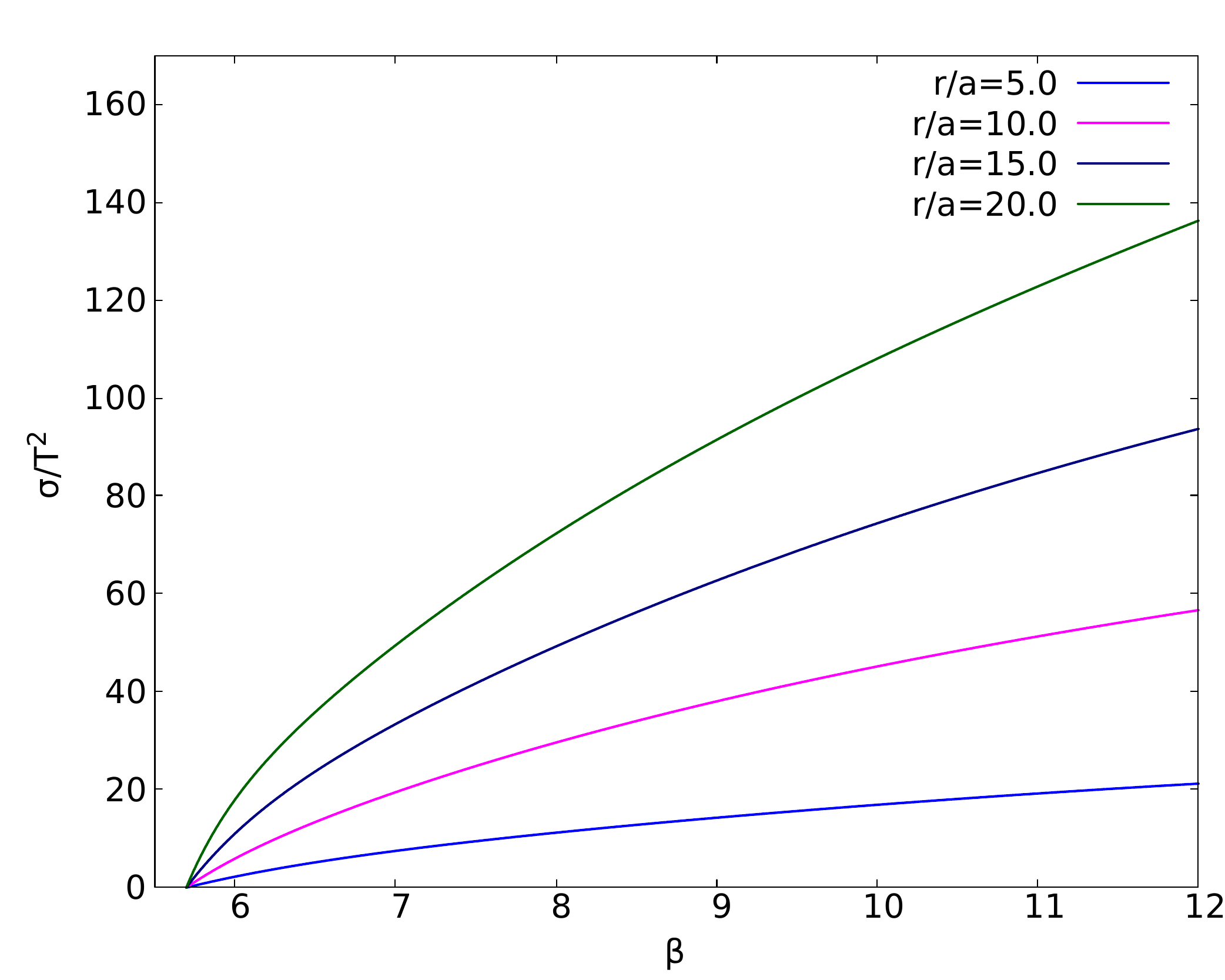}
\caption{For $N_\tau=4$, $\beta$ dependence of $\sigma/T^2$ plot for different radial patches $r/a=5,10,15,20$.}
\label{fig:sigmaT_beta_nt4}
\end{figure}

Integrating the fitted function $f(\beta)$ yields the free energy of
the string configuration and the corresponding string tension. The
dependence of $\sigma/T^2$ on $\beta$ shown in Fig.\ref{fig:sigmaT_beta_nt4}. 
We see that the free energy for the string as well as the domain
walls rises slower near $\beta_c$
as compared to $N_\tau=2$, with a rapid rise near $\beta_c$ and an
approximately linear increase at larger $\beta$. These results indicate
that the qualitative features of the string configurations remain
unchanged  with the lattice cut-off.

For both $N_\tau=2$ and $N_\tau=4$, the rise in $\sigma/T^2$ is larger
for bigger radial patches. For smaller patches the contribution of the
string core dominates the free energy, while for larger patches the
contribution from the domain walls becomes dominant as they extend to
the boundary of the system. These observations indicate that the free
energy contribution from the core grows more slowly with $\beta$
compared to the contribution from the domain walls. Moreover for larger
patches the string tension and interface tension results agree up to an
overall $\beta$ independent scale factor, confirming that the interface
tension dominates the string tension.

\section{Conclusion}

In this work, we have conducted a detailed first-principles non-perturbative study of string configurations in the deconfined phase of $SU(3)$ gauge theory, originating from the spontaneous breaking of $Z_3$ symmetry. Using the Polyakov loop effective potential model, we argued that these strings are topologically stable. In the lattice simulations, our focus is mainly on the winding one string, which results from the junction of three domain walls. We used specific boundary conditions that lead to the formation of a string with three domain walls connected. Our methodology, based on indirect free energy estimation through action differences, was validated by recovering known results for the $Z_3$ interface tension, in qualitatively good agreement with previous studies. This suggests that our approach reliably captures the essential physics of the string configurations.  Subsequently, we calculated the string tension, $\sigma/T^2$ as a function of $\beta$. Our results show that the string tension rises steeply near $\beta_c$. For large $\beta$(weak coupling limit) the string tension for $N_\tau=2$ and 
$N_\tau=4$ approaches $\sqrt{\beta}$ for large patches as the domain walls dominate
over the string core. The $Z_3$ string being a global defect, the string tension rises with the size of the area of the cross section. For a $U(1)$ global string, the tension approaches $log(r)$, where $r$ is the radius of the cross section. In this present case, the string tension for larger radius is dominated by the interface tension, hence rises linearly in r. 

Our simulations neglect the effects of dynamical quarks, whose explicit breaking of $Z_3$ symmetry would likely make these configurations non-static. The explicit breaking will make the strings unstable near the critical temperature. However, for temperatures far above $T_c$ they are expected to form and affect the dynamics of the system. In future, we plan to study the effect of $Z_3$ explicit breaking on these configurations by incorporating dynamical fermions.

\begin{acknowledgments}
We would like to thank Ajit M. Srivastava for his valuable comments and suggestions.
\end{acknowledgments}

\end{document}